\documentclass[a4paper,11pt]{article}
\usepackage{graphicx}
\usepackage{fullpage}
\usepackage{setspace}
\usepackage{amsmath}
\usepackage{amssymb}
\usepackage{url}
\usepackage{bm}

\pdfoutput=1
\usepackage{hyperref}
\hypersetup{
  pdfinfo={
    Title={Synthesized multiwall MoS2 nanotube and nanoribbon 
    field-effect transistors},
    Author={S. Fathipour}
  }
}
\usepackage{pdfpages}

\begin{document}


# Synthesized multiwall MoS$_2$ nanotube and nanoribbon field-effect transistors

S. Fathipour,[1,a)] M. Remskar,[2] A. Varlec,[2] A. Ajoy,[1] R. Yan,[3] S. Vishwanath,[1] W. S. Hwang,[4] H. G. Xing,[1] D. Jena,[1] and A. Seabaugh[1,b)]

[1] Department of Electrical Engineering, University of Notre Dame, Notre Dame, Indiana 46556, USA
[2] Solid State Physics Department, Jožef Stefan Institute, Ljubljana, Slovenia
[3] Department of Electrical Engineering, Cornell University, Ithaca, New York 14850, USA
[4] Department of Materials Engineering, Korea Aerospace University, Gyeonggi 412791, Korea

We report on the fabrication and characterization of synthesized multiwall MoS$_2$ nanotube (NT) and nanoribbon (NR) field-effect transistors (FETs). The MoS$_2$ NTs and NRs were grown by chemical transport, using iodine as a transport agent. Raman spectroscopy confirms the material as unambiguously MoS$_2$ in NT, NR, and flake forms. Transmission electron microscopy was used to observe cross sections of the devices after electrical measurements and these were used in the interpretation of the electrical measurements allowing estimation of the current density. The NT and NR FETs demonstrate *n*-type behavior, with ON/OFF current ratios exceeding $10^3$, and with current densities of 1.02 µA/µm, and 0.79 µA/µm at $V_{DS}$ = 0.3 V and $V_{BG}$ = 1 V, respectively. Photocurrent measurements conducted on a MoS$_2$ NT FET, revealed short-circuit photocurrent of tens of nanoamps under an excitation optical power of 78 $\mu$W and 488 nm wavelength, which corresponds to a responsivity of 460 $\mu$A/W. A long channel transistor model was used to model the common-source characteristics of MoS$_2$ NT and NR FETs and was shown to be consistent with the measured data.

---

a) Electronic mail: sfathipo@nd.edu
b) Electronic mail: seabaugh.1@nd.edu

There is a growing interest in atom-thick materials such as graphene and the transition metal dichalcogenides (TMDs) for electronics applications[1,2]. Bulk and single layer sheets of TMDs are beginning to be explored in a wide range of optoelectronic and electronic devices including solar cells[3,4], photodetectors[5,6], sensors[7,8], field-effect transistors[9,10,11,12], and logic circuits[13,14]. The absence of surface dangling bonds, the excellent gate electrostatics of the few-layer transistor, and the potential for large area planar processing are all motivating this research. While planar processing is desirable for manufacturing, at the limits of scaling, the properties of these materials may be compromised by unpassivated dangling bonds at the sheet edges. Edges introduce traps, which can degrade subthreshold swing and increase tunneling leakage, $1/f$ noise, and variability in the device characteristics. Edges can be substantially eliminated by using nanotubes (NTs), and nanoribbons (NRs) formed from collapsed NTs.

There are only a few experimental reports of TMD NT transistors and these are based on $WS_2$[15,16] and $MoS_2$[17,18]. Density functional theory shows that $WS_2$ and $MoS_2$ NTs are semiconducting with well-defined band gaps[19,20]. Levi[15] reported *n*-channel conduction on a 75 nm diameter $WS_2$ multiwall NTs with an ON/OFF current ratio of about 2; this commercial material was dispersed from powder. Unalan[16] used multiwall NTs synthesized by conversion of tungsten oxide particles[21]; the ON/OFF current ratio was about 4. The first $MoS_2$ NT FET report by Remškar[17] showed no gate modulation and no photoresponse on a multiwall NT with a diameter of approximately 100 nm and synthesized from $Mo_6S_4I_6$ nanowires by sulphurization using $H_2S/ H_2/$ Ar gas mixture, the method that causes a high density of structural defects[22]. In a recent report, Strojnik[18] confirmed transistor action with *n*-channel conduction in $MoS_2$ NTs based on the same synthesis method and achieved an ON/OFF current ratio of 60. In this paper we report on the properties of $MoS_2$ NT and NR FETs in which an ON/OFF current ratio exceeding $10^3$ is achieved.

The NTs used in this study were grown by chemical transport reaction, which enables growth of NTs with very slow rate from the vapor phase[23]. The silica ampoule containing $MoS_2$ powder and iodine in amount of 1.5 mg/cm$^3$ was evacuated and sealed at a pressure of 7 x $10^{-4}$ Pa. The transport reaction using iodine as a transport agent ran from 1133 K to 1010 K with a temperature gradient of 6.2 K/cm in a two-zone furnace. After three weeks of growth, the silica ampoules were cooled to room temperature with a controlled cooling rate of 60º C/hour. Approximately a few percent of the starting material was transported by the reaction to form nanotubes, while the rest of the transported material grows as strongly undulated thin plate-like crystals. The nearly equilibrium growth conditions enable the synthesis of nanotubes of different diameters, length, and wall thickness, but with extremely low density of structural defects. They grow up to several millimeters in length. The

diameters in multi-wall nanotubes range from several micrometers to less than ten nanometers. Some of the nanotubes spontaneously collapse during the growth and continue in growth in ribbon shapes[24].

The completed transistor planar views and schematic cross section are shown in Figure 1. The process flow consisted of electron-beam evaporation of Ti/Au (5/100 nm) on the back of the Si wafer. The $MoS_2$ nanostructures were exfoliated using 3M Scotch 810 tape onto 26 nm $Al_2O_3$ formed by atomic layer deposition. The NTs and NRs were then patterned to form the source and drain contacts using electron beam lithography followed by deposition of Sc/Ni (40 nm/ 20 nm) metal contacts. Figure 1 shows (a) scanning electron microscope (SEM) image, (b) atomic force microscope (AFM) image, and (c) schematic cross section of the FET.

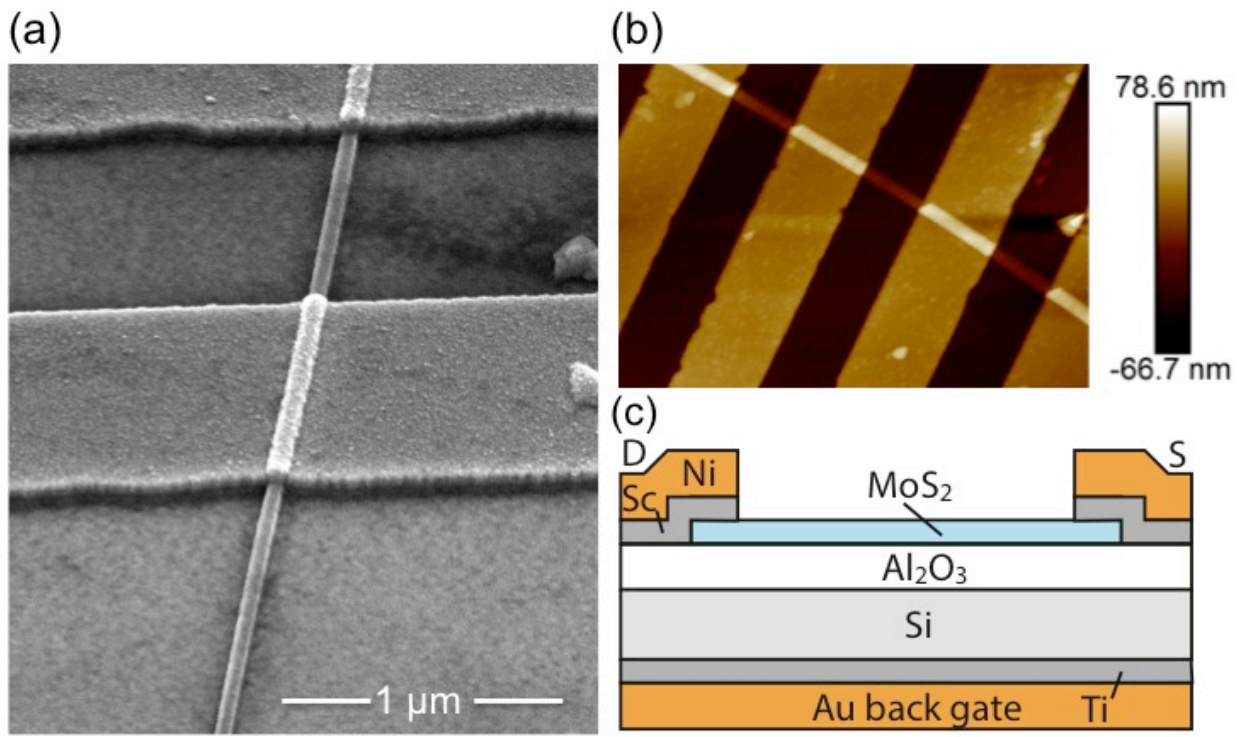

FIG. 1. (a) SEM image, (b) AFM image and (c) schematic cross-section of a $MoS_2$ NT FET. The four-contacts Kelvin connection shown in (b) was used to measure the contact resistance.

Transmission electron microscope (TEM) images were taken following the electrical measurements to establish clearly the cross section of the measured devices. Fig. 2, images (a) and (b) are of the NR and (c) shows the NT. Both nanostructures are multiwall and the ribbon is a collapsed nanotube with a total thickness of about 7 nm and 10 monolayers. This would give an interlayer distance of 0.7 nm for the NR. As shown by the TEM image of Fig. 2(c), NT has an elliptical shape with a wall thickness of 11 layers. Interlayer distance in NT based on the TEM results is 0.68 nm. The interlayer distance in both NR and

3

NT, is close to the range of 0.62 to 0.68, reported in literature[25,26,27]. High-resolution TEM images, not shown, reveal that the MoS$_2$ has the 2H hexagonal structure.

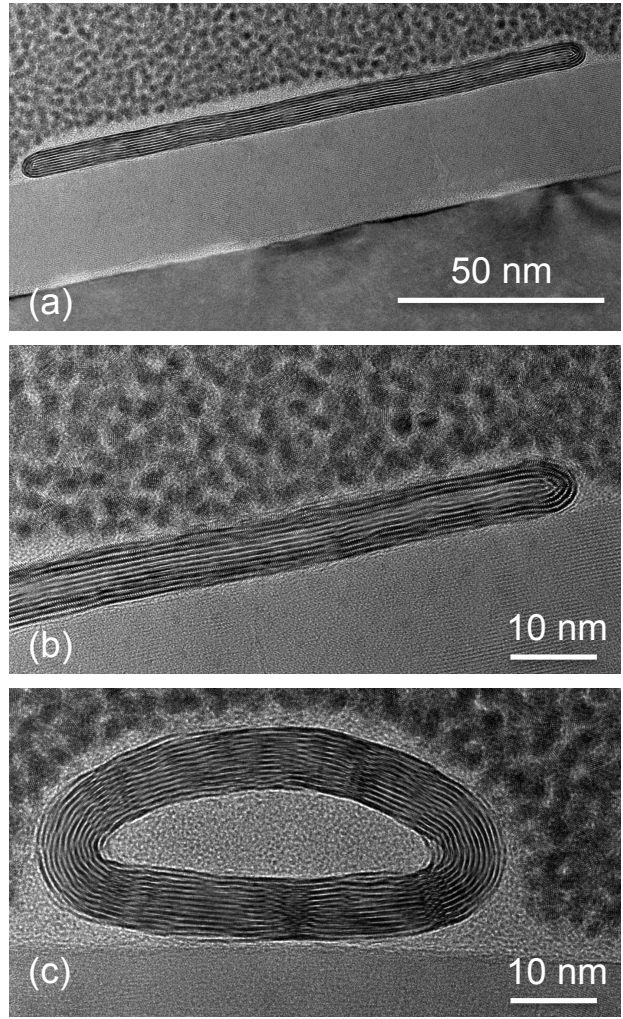

FIG. 2. Transmission electron microscope image of the transistor channel cross sections: (a) MoS$_2$ nanoribbon with width of 140 nm, height of 7 nm, and cross sectional area of 980 nm$^2$. The nanoribbon has a layer thickness of 10 monolayers. (b) MoS$_2$ nanoribbon magnified to show wrapping of the layers at the ribbon edge. (c) MoS$_2$ NT with a layer thickness of 11 monolayers and cross sectional area of 766 nm$^2$. The minor and major radii for the inner ellipse are 5 nm and 20 nm, respectively, and for the outer ellipse are 12.5 nm and 27.5 nm.

Raman measurements, shown in Fig. 3(a), were performed on the NR, the NT, and a flake that was exfoliated from the same material source. Measurements were done in the backscattering configuration using a WITec Alpha 300 system at room temperature (100× objective, 488 nm laser wavelength, 144 kW/cm$^2$ power density). Clear signals of Raman in-plane vibrational mode, $E_{2g}^1$, and out-of-plane vibrational mode, $A_{1g}$, for MoS$_2$ are observed in all the grown nanostructures. A slight red shift is observed in the Raman peaks of the NT and NR structures relative to the flake. Local heating due to laser illumination is known to red shift both the $E_{2g}^1$ and $A_{1g}$ peaks in bulk, and few-layered MoS$_2$ as well as in MoS$_2$

microtubes[28,29]. The conduction of heat out of the NT may be expected to be less than the NR because of smaller contact area to the substrate. Thus, heat may be expected to red shift the NT more than the NR, which is what is observed.

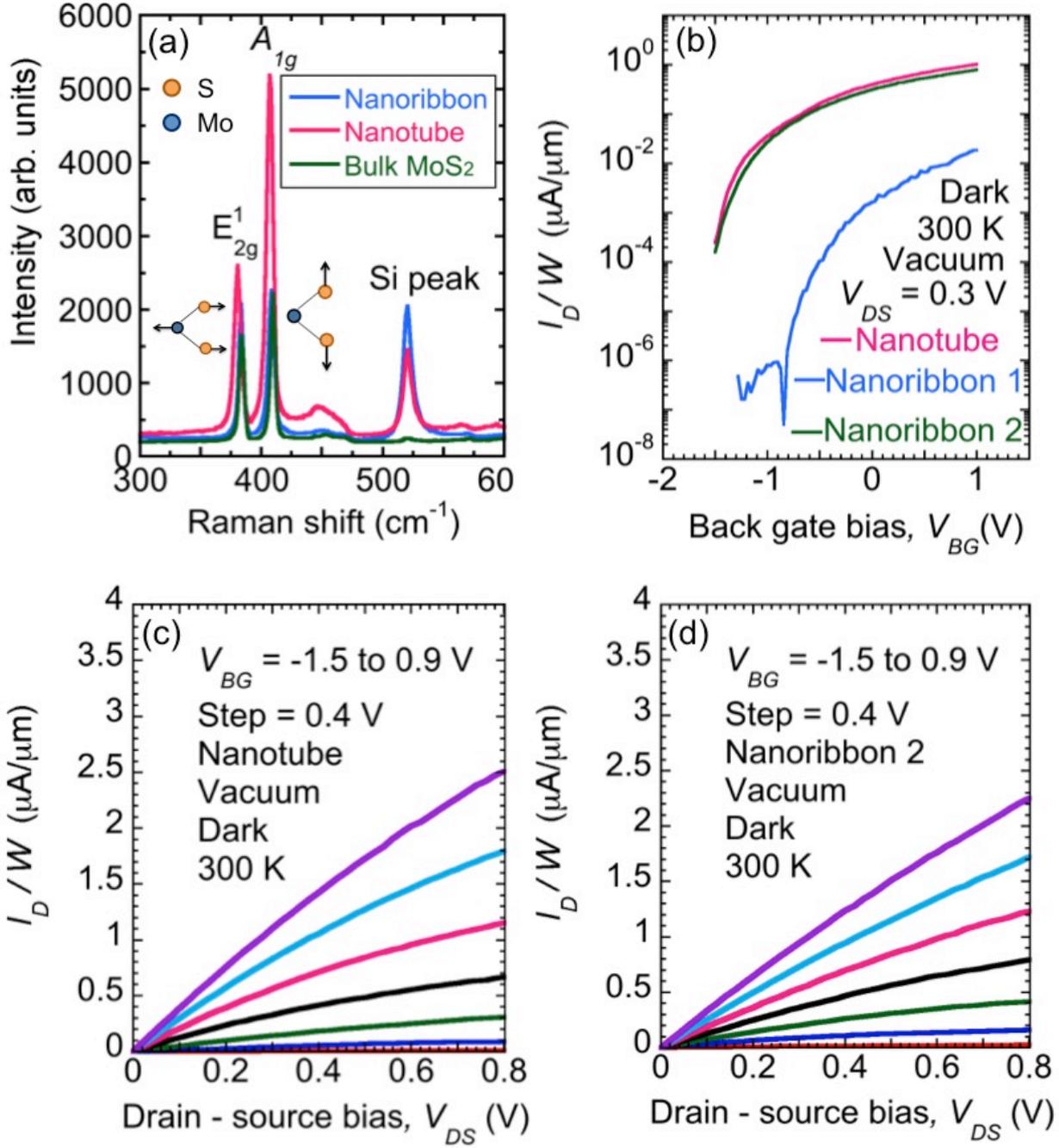

FIG. 3. (a) Raman spectrum of the $MoS_2$ NT, NR and bulk $MoS_2$ flake with a laser wavelength of 488 nm and power density of 144 kW/cm$^2$. The out-of-plane $A_{1g}$ and in-plane $E_{2g}^1$ vibrations are identified in the inset. (b) Measured drain current per unit width vs. back-gate voltage for the $MoS_2$ NT FET and two $MoS_2$ NR FETs, designated NR1 and NR2. Channel widths are 55 nm, 140 nm, and 321 nm, respectively. (c) Common source characteristics for the $MoS_2$ NT FET measured in vacuum. (d) Common source characteristics for the $MoS_2$ NR2 FET measured in vacuum.

A smaller $E_{2g}^1$ over $A_{1g}$ ratio was observed in the NT. This could be explained by the fact that in the back-scattering configuration, polarization of incoming light is perpendicular to the c-axis of the crystal in the NR, bulk MoS$_2$ and center of NT, while it is parallel to the c-axis at the edges of the NT. Direction of the $A_{1g}$ mode vibration is parallel to the c-axis[30] and therefore its intensity increases at nanotube edges. The edges can contribute majority of the signal when diameter of NT is as small as tens of nanometers.

The measured drain current per unit width vs. back-gate voltage is shown in Fig. 3(b) for a NT and two NR FETs, designated NR1 and NR2. All transistors showed *n*-channel conduction, with ON/OFF current ratios of more than $10^3$. In the course of the measurement, the reverse bias was limited to −1.5 V, wherein the leakage current was less than a few pA. As a result, the value of $10^3$ for the NT and NR2 FET should be considered a lower limit. This value for the NT FET far exceeds the prior best report of 60 by Strojnik[18]. From the measured dimensions given in the caption of Fig. 2, the width for the NT and the NR1 can be estimated to be 55 nm and 140 nm. The measured width for NR2 is 321 nm. The corresponding values for the extracted current per unit width measured in NT, NR1 and NR2 FETs are 1.02 µA/µm, 0.02 µA/µm, and 0.79 µA/µm at $V_{DS}$ = 0.3 V and $V_{BG}$ = 1 V. The lower current density in NR1 FET is not well explained by contacts or mobility and remains a question.

The electrical characteristics of the back-gated MoS$_2$ NT and NR2 FETs are further modeled using a long channel FET model. The detailed description of this model is included in the supplementary material[31]. Carrier density, mobility and flat band voltage are the unknowns in the model. Using a charge density of 1x10$^{16}$ cm$^{-3}$, mobility of 43.5 cm$^2$/ Vs and flat band voltage of -0.9 V for the MoS$_2$ NT FET and a charge density of 1x10$^{16}$ cm$^{-3}$, mobility of 36 cm$^2$/ Vs and flat band voltage of -1 V for the MoS$_2$ NR2 FET, an excellent fit to the experimental data was obtained, see supplemental material.

The sole report of MoS$_2$ NR FETs, by Liu[32], utilized reactive ion etching to form ribbons as narrow as 60 nm from exfoliated flakes. Liu achieved an ON/OFF current ratio of nearly $10^4$ on n-channel FETs for a 60 nm NR width with a 6 nm thickness. This exceeds the $10^3$ ON/OFF ratio obtained here. Subthreshold swing is much improved in this paper, Our minimum subthreshold swing for both NR FETs is 200 mV/decade, compared to 10 V/decade in the MoS$_2$ NR FET by Liu[32]. The enhanced subthreshold swings in the NR can be attributed to two factors. Firstly, the NR in our FET is grown. Therefore, edge roughness and defects due to dangling bonds are largely nonexistent. This is in contrast with the NRs formed by plasma

dry etching of $MoS_2$ flakes by Liu. Secondly, to fabricate devices, we use 26 nm of $Al_2O_3$, whereas Liu uses 300 nm $SiO_2$. This will extend the SS by factor of 10×, bringing the two results more closely in line.

Four point probe measurements were used to extract the contact resistance of the $MoS_2$ NT FET with a Sc contact. The extracted value for contact resistance at $V_{BG}$ = 0 V was 39.6 kΩ.μm. This exceeds the reported value of 10 kΩ.μm by Das in [33] for same overdrive voltage and with same contact on $MoS_2$. As outlined by Das[33], the extracted resistance involves contributions from both Schottky barrier resistance and also resistive coupling between the layers of the 2D material. Therefore, the resistance due to the Schottky barriers may be less than this value.

Scanning photocurrent measurements were also conducted on $MoS_2$ NT FETs, fabricated with Ti/ Au (1 nm/100 nm) source and drain contacts on a highly-doped Si wafer covered by 285 nm $SiO_2$. A WITec Alpha 300 scanning confocal microscope was used to focus a 488 nm wavelength laser onto the NT FET. The beam diameter was determined to be about 660 nm. Using spot area and the measured laser power, a power density of 22 kW/cm$^2$ was determined. Photocurrent was recorded as the laser spatially scanned the sample. Current was converted to a voltage using a transimpedance amplifier. The reflected light was simultaneously collected to correlate with the spatial photocurrent mapping. Fig. 4(a), shows a representative short-circuit photocurrent ($V_{DS}$ = $V_{BG}$ = 0) mapping of a $MoS_2$ NT FET, with the metal 1 as drain, metal 2 floating, and metal 3 as source. Fig. 4 (b) shows the photocurrent profile along the blue dotted line of Fig. 4(a). Prominent negative and positive photocurrents are detected at the drain and source side, respectively. The photocurrent is due to the separation of photogenerated carriers by band bending at the metal-semiconductor junctions, which dominate the voltage drop in the TMD FET. Responsivity is defined as the ratio of photocurrent, divided by the incident power on the nanotube. The incident power on the nanotube was found by multiplying the total power of the laser and the area ratio of the nanotube and laser spot. The responsivity is ~ 460 $\mu$A/W close to that of graphene reported in [34].

Using the Schottky photocurrent mapping of Fig. 4(b), we estimate the band bending profile near source/ $MoS_2$ junction and drain/ $MoS_2$ junction. When the NT is illuminated near the drain or source contact, electron/ hole pairs are locally generated at the illumination region. Under the built in Schottky field at NT/metal junction, the electron and the hole are separated with the current direction set by the band bending. To match the observed photocurrent polarity, the bands are bent upward, as indicated in the band diagram shown in the inset of Fig. 4(b).

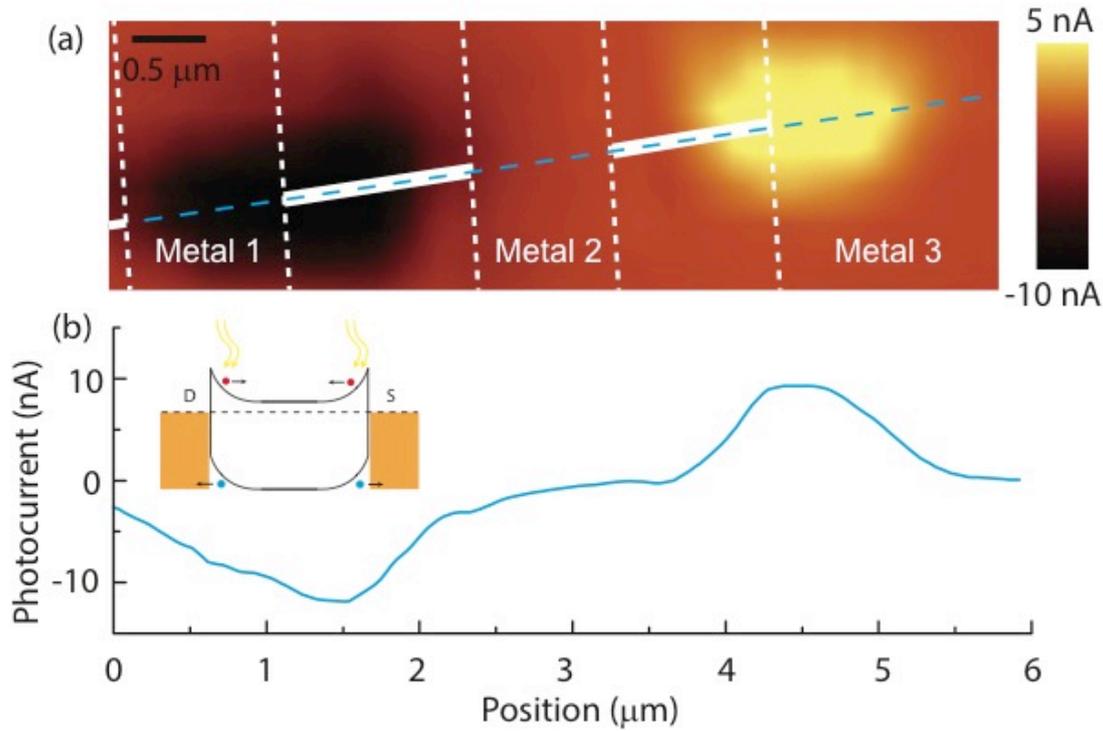

FIG. 4. (a) Representative short-circuit photocurrent mapping of a $MoS_2$ NT FET. (b) The line profile is the photocurrent along the dotted blue line in (a). The inset shows the schematic of the band bending in the $MoS_2$ NT device. Blue and red carriers refer to holes and electrons, respectively.

In conclusion, synthesized $MoS_2$ NT and NR FETs were fabricated and characterized. Devices demonstrated *n*-type characteristics with ON/OFF current ratios of more than $10^3$, greatly exceeding the best prior report of 60 in the NT case. Raman measurements confirm the $E_{2g}^1$ and $A_{1g}$ vibrational modes in both tubes and ribbons with a red shift compared to bulk $MoS_2$, due to laser heating effect. Photocurrent measurements were conducted on $MoS_2$ NT FET with Ti/ Au contacts and revealed a responsivity of ~ 460 $\mu$A/W. The common-source characteristics were shown to be in good agreement with a long channel transistor model.

**ACKNOWLEDGMENTS**

8


\newpage

\begin{center}
 { \Large {\textbf{ {Supplemental Material \\}}} }
 { \large {\textbf{Synthesized multiwall MoS$_2$ nanotube and nanoribbon field-effect transistors \\}} }
 { S. Fathipour $^{1}$, M. Remskar $^2$, A. Varlec $^2$, A. Ajoy $^{1,a}$, R. Yan $^3$, S. Vishwanath $^1$, W. S. Hwang $^4$,
H. G. Xing $^1$, D. Jena $^{1,b}$, and A. Seabaugh $^{1}$\\}
\end{center}

\begin{small}
\noindent
$^1$  Department of Electrical Engineering, University of Notre Dame, Notre Dame, Indiana 46556, USA \\
$^2$  Solid State Physics Department, Jo\v{z}ef Stefan Institute, Ljubljana, Slovenia \\
$^3$  Department of Electrical Engineering, Cornell University, Ithaca, New York 14850, USA \\
$^4$  Department of Materials Engineering, Korea Aerospace University, Gyeonggi 412791, Korea \\
\end{small}

\begin{small}
\noindent
a) \url{aajoy@nd.edu} \\
b) \url{djena@nd.edu} \\
\end{small}
 
\begin{figure}[b!]
\centering
\includegraphics{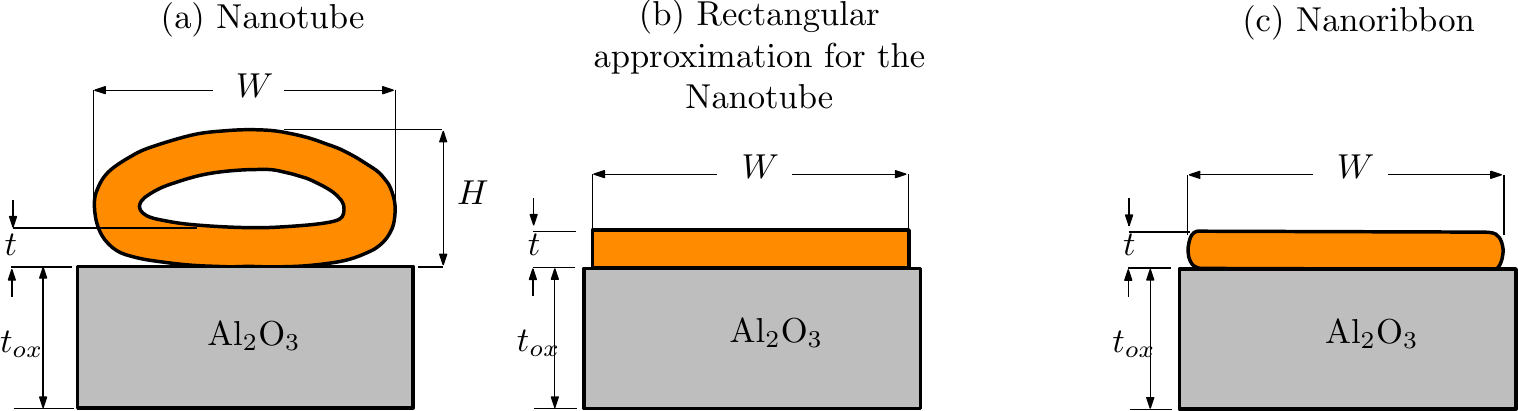}
\caption{(a),  (c) Cross-sections  of  the NT/NR  MOSFETs along  their
width,  and  a  rectangular   approximation  for  the  NT  (b).   This
approximation is reasonable when the channel is strongly acccumulated,
since the  channel lies close to the  oxide-semiconductor interface in
this regime of operation.}
\label{fig_approximation}
\end{figure}

In this supplemental material, we attempt to model the current-voltage
characteristics  of  the MoS$_2$  nanotube  (NT)  and nanoribbon  (NR)
MOSFETs. The  growth of the NTs  and NRs results in  the MoS$_2$ being
unintentionally $n$-doped.  The  fabricated $n$-type MOSFETs are hence
accumulation-depletion devices -- the  channel is accumulated when the
device is in  the on-state, and is depleted of  carriers as the device
turns off. In accumulation, the channel is dominated by electrons that
are close  to the oxide-semiconductor  interface. As the  device turns
off, the channel is dominated  by electrons that are farther away from
this interface. We  restrict ourselves to gate voltages  for which the
channel is  accumulated.  In  this regime, it  is hence  reasonable to
approximate the  cross section  of the NT  MOSFETs by a  rectangle, as
shown  in  Fig.    \ref{fig_approximation}.   The  thickness  of  this
rectangle corresponds  to the thickness of  the wall of  the NT.  This
approximation  allows us  to  apply results  derived  for an  $n$-type
junctionless  FET \cite{Chen_TED_2012}  to model  both the  NT  and NR
MOSFETs.

\newpage

\begin{figure}[t]
\centering
\includegraphics{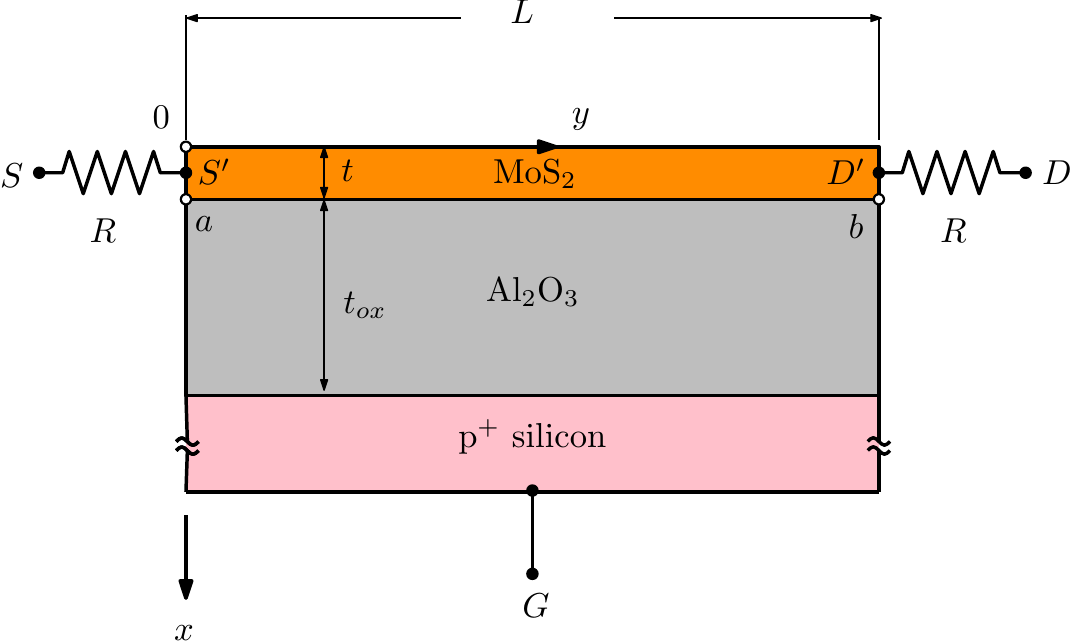}
\caption{Cross-section of the  approximate NT (Fig. \ref{fig_approximation}(b))
and NR (Fig. \ref{fig_approximation}(c)) devices  along their length.}
\label{fig_paosah}
\end{figure}

\section{Model}
\begin{figure}[t!]
\centering
\includegraphics{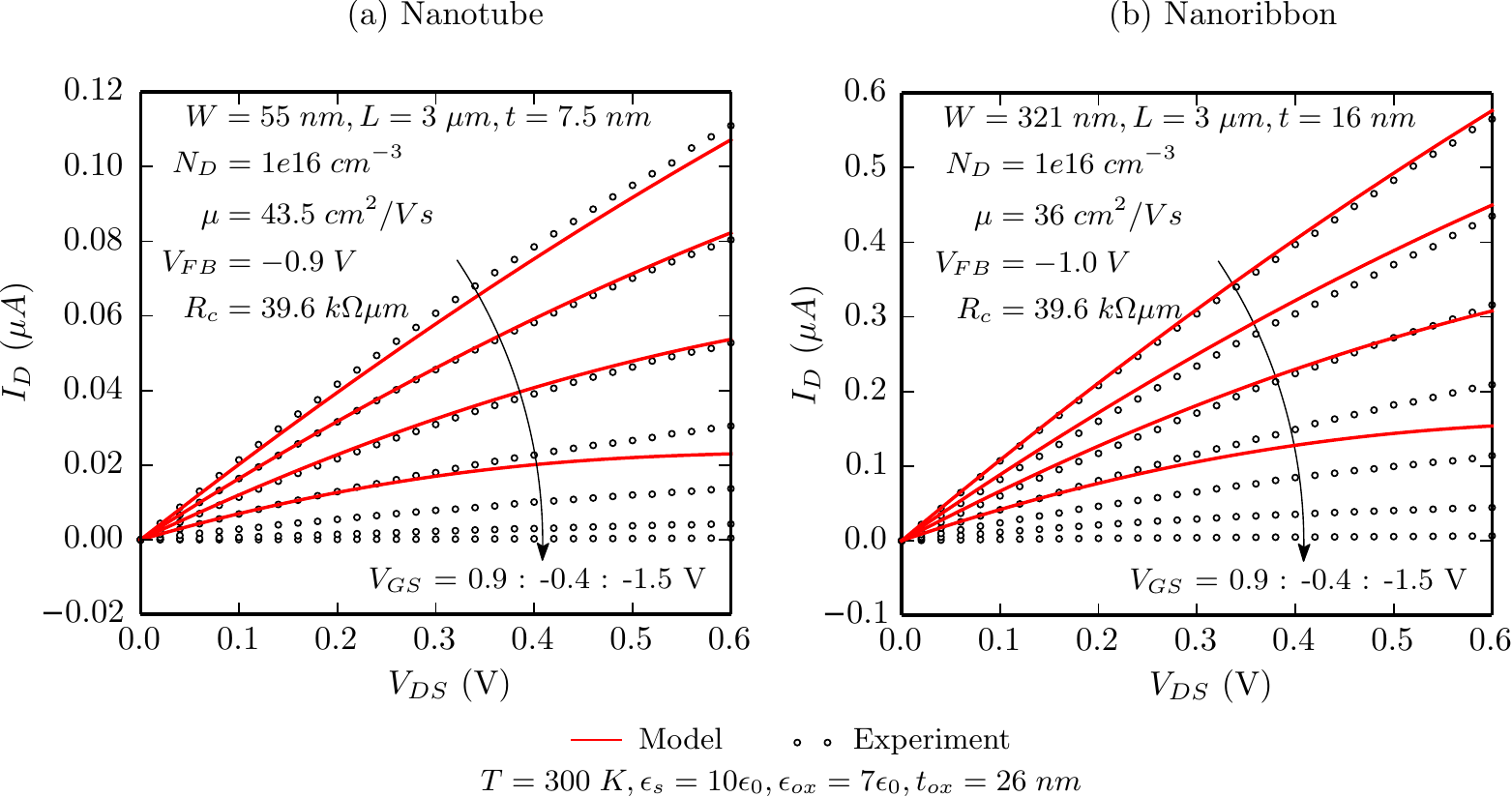}
\caption{Comparison of  experimental data with results  from the model
for the (a) nanotube and (b) nanoribbon MOSFETs. Parameter values used
for the simulation are also shown. $W, L, t, t_{ox}, T$ for the NT and
NR, $R_c = R \cdot W$ for the NT are measured values. We have used the
same  value  of $R_c$  for  the NR  too.  Values  of $\epsilon_s$  and
$\epsilon_{ox}$  are  taken  from  \cite{Kim_NatureComm_2012}.   Other
parameters have been obtained by fitting the data to the model.}
\label{fig_devicechar}
\end{figure}

See Fig. \ref{fig_paosah}. $S$, $D$  and $G$ refer to the external terminals of
the MOSFET. Source and  drain contact resistances (assumed equal, $R$)
connect $S$, $D$ to intrinsic  nodes $S'$ and $D'$ respectively.  When
a current $I_D$  flows through the device, the  terminal and intrinsic
voltages are given by $V_{GS'} = V_{GS} - I_D R$, $V_{D'S'} = V_{DS} -
2I_DR$.  Within  the intrinsic  MOSFET, the electron  concentration is
$n(x,y)  =  N_D  \exp{( (  \phi(x,y)  -  V(y)  )  /  V_t )  }$,  where
$\phi(x,y)$ is the quasi-Fermi  potential, $V(y)$ is the electrostatic
potential   \cite{Pao_SSE_1966,    Ortiz_SSE_1992},   $N_D$   is   the
unintentional   doping  concentration,  and   $V_t$  is   the  thermal
voltage. The surface  potential $\phi_s(y) \equiv \phi( t,  y)$ at the
oxide-semiconductor  interface  is  obtained  by  a  solution  of  the
following implicit equation
\begin{align}
( V_{GS'} - V_{FB} - \phi_s )^2 = \frac{ 2 \epsilon_s q N_D V_t }{C_{ox}^2}
          \left[ \exp{ \left( \frac{\phi_s - V}{V_t} \right)} - 1 \right] 
\label{eq_surfacepot}
\end{align}
where  $V_{FB}$   is  the  flat-band  voltage,   $\epsilon_s$  is  the
dielectric  constant of  MoS$_2$ and  $C_{ox}  = \epsilon_{ox}/t_{ox}$
with  $\epsilon_{ox}$,  $t_{ox}$  being  the dielectric  constant  and
thickness of the Al$_2$O$_3$ respectively. Note that this equation has
been  derived under  the condition  that the  net charge density at  the rear
interface  ($x =  0$) is  zero  in accumulation (which corresponds  to
$\phi(0,y)  = V(y)$).  Setting  $V(0) =  0$ and  $V(L) =  V_{D'S'}$ in
eq. (\ref{eq_surfacepot})   respectively  yields   the   surface  potentials
$\phi_a$  and $\phi_b$  (corresponding to  the points  $a,b$  shown in
Fig. \ref{fig_paosah}). The current is then
\begin{align}
\label{eq_current}
I_D = \mu C_{ox} \frac{W}{L} \Bigg[ & 2 V_t \phi_s +
      2 V_t \sqrt{ \beta V_t } \arctan{ \left( \frac{V_{GS'} - V_{FB} - \phi_s}
                                               {\sqrt{ \beta V_t} } \right) } \\
\nonumber    &  - \frac{1}{2} ( V_{GS'} - V_{FB} - \phi_s)^2 \Bigg]_{\phi_a}^{\phi_b} +
      \mu \frac{W}{L} q N_D t  V_{D'S'}
\end{align}
where  $\mu$  is  an effective  value  of  mobility,  and $\beta  =  2
\epsilon_s q  N_D / C_{ox}^2$.  A comparison of the  experimental data
with    the   results    of   the    above   model    is    shown   in
Fig. \ref{fig_devicechar}.

\section{Verification of zero charge density approximation}

\begin{figure}[t!]
\centering
\includegraphics{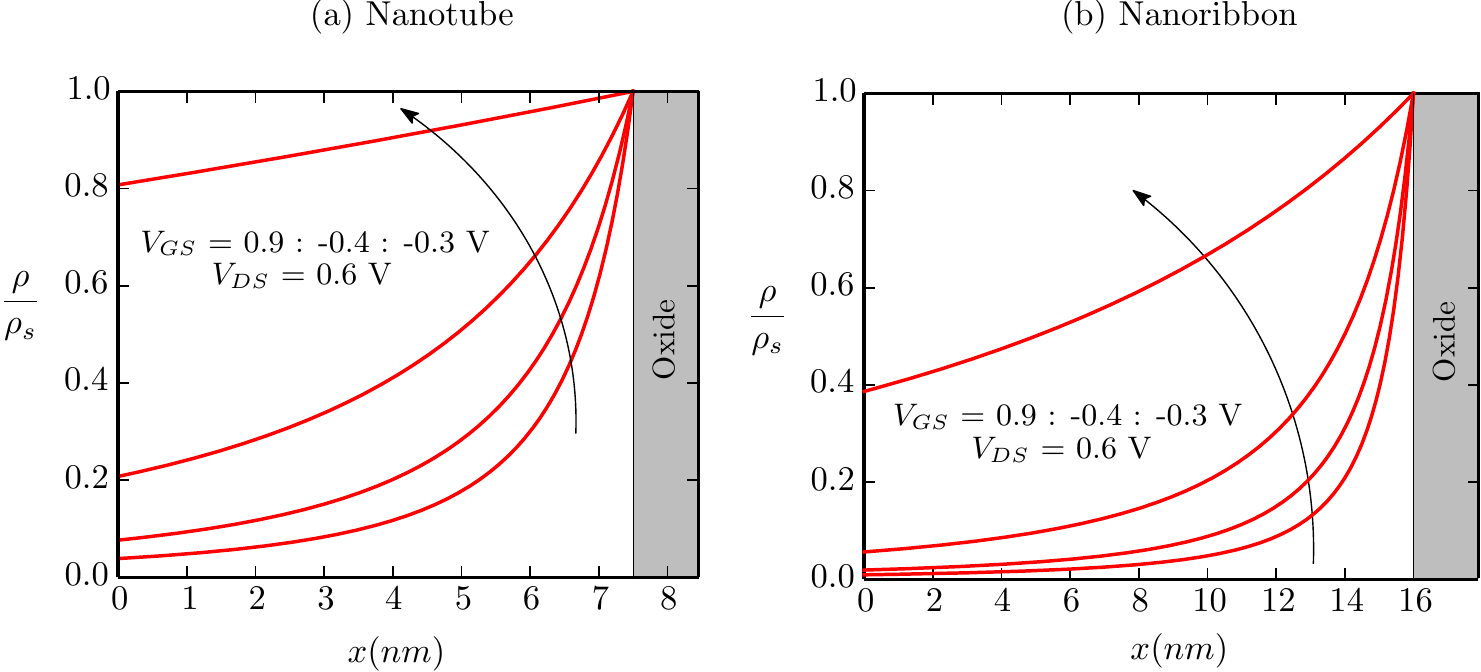}
\caption{Verification of the approximation  that charge density at the
rear interface  $\rho(0) = 0$  using a solution of  eq. (\ref{eq_ode})
for  the (a)  nanotube and  (b) nanoribbon  mosfets.  $\rho_s$  is the
charge density at the MoS$_2$-oxide  interface $(x = t)$.  Notice that
the  difference  between  the  model  and experimental  data  in  Fig.
\ref{fig_devicechar} is largest where  the approximation $\rho(0) = 0$
is the worst.}
\label{fig_chargesheet}
\end{figure}

We  verify whether  the  approximation  of  zero net  charge
density  at  $x=0$  is  valid  for  the  $I_D$-$V_D$  curves  we  have
modeled. To do this, we solve the electrostatic problem \emph{without}
this  approximation, in terms  of the  surface potential  $\phi_s$ and
surface electric  field $E_s$ obtained  using the zero  charge density
approximation. This allows us to  plot the charge density $\rho(x)$ as
a function  of $x$.  Our approximation is  reasonable if  $\rho(0)$ so
obtained is close to zero.  We begin with Poisson's equation under the
gradual channel approximation given by
\begin{align}
\label{eq_poisson}
\frac{d^2 \phi}{dx^2} = \frac{-\rho}{\epsilon_s} = 
     \frac{q N_D}{\epsilon_s} \left( \exp {\left( \frac{\phi - V }{V_t} \right)} - 1\right)
\end{align}
Integrating eq. (\ref{eq_poisson}) by parts,  we get $\int d(E^2) = -2
\int ( \rho(  \phi ) / \epsilon_s ) d\phi $  (where the electric field
$E = -d \phi / dx$), which yields
\begin{align}
\label{eq_ode}
E = -\sqrt{E_s^2 - \frac{2 q N_D V_t}{\epsilon_S} \left[ \exp{ \left( \frac{\phi_s - V}{V_t}\right) }
          - \exp{ \left( \frac{\phi - V}{V_t}\right) } 
          - \exp{ \left( \frac{\phi_s - \phi}{V_t}\right) } \right ]}
\end{align}
The negative sign for the  square root is sensible in the accumulation
regime.  Eq.   (\ref{eq_ode}) can be cast as  an ordinary differential
equation $  d\phi/dx' =  f( \phi )$,  with $x'  = t -  x$, and  can be
solved numerically to obtain $\phi(x)$ and $\rho(x)$, given the values
of   $E_s,  \phi_s$   (which  were   obtained  from   a   solution  of
eq. (\ref{eq_surfacepot}) ). The results from this calculation are shown
in Fig. \ref{fig_chargesheet}. The value of $V$ in eq. (\ref{eq_ode}) is
chosen to be equal to $\max(V_{D'S'})$. Notice that the approximation 
$\rho(0) \approx 0$ is the best for the largest value of $V_{GS}$, and worst
for the lowest value of $V_{DS}$. This corresponds to the trends seen in
Fig. \ref{fig_devicechar}.

\section{Projection assuming ideal contacts}
\begin{figure}[t!]
\centering
\includegraphics{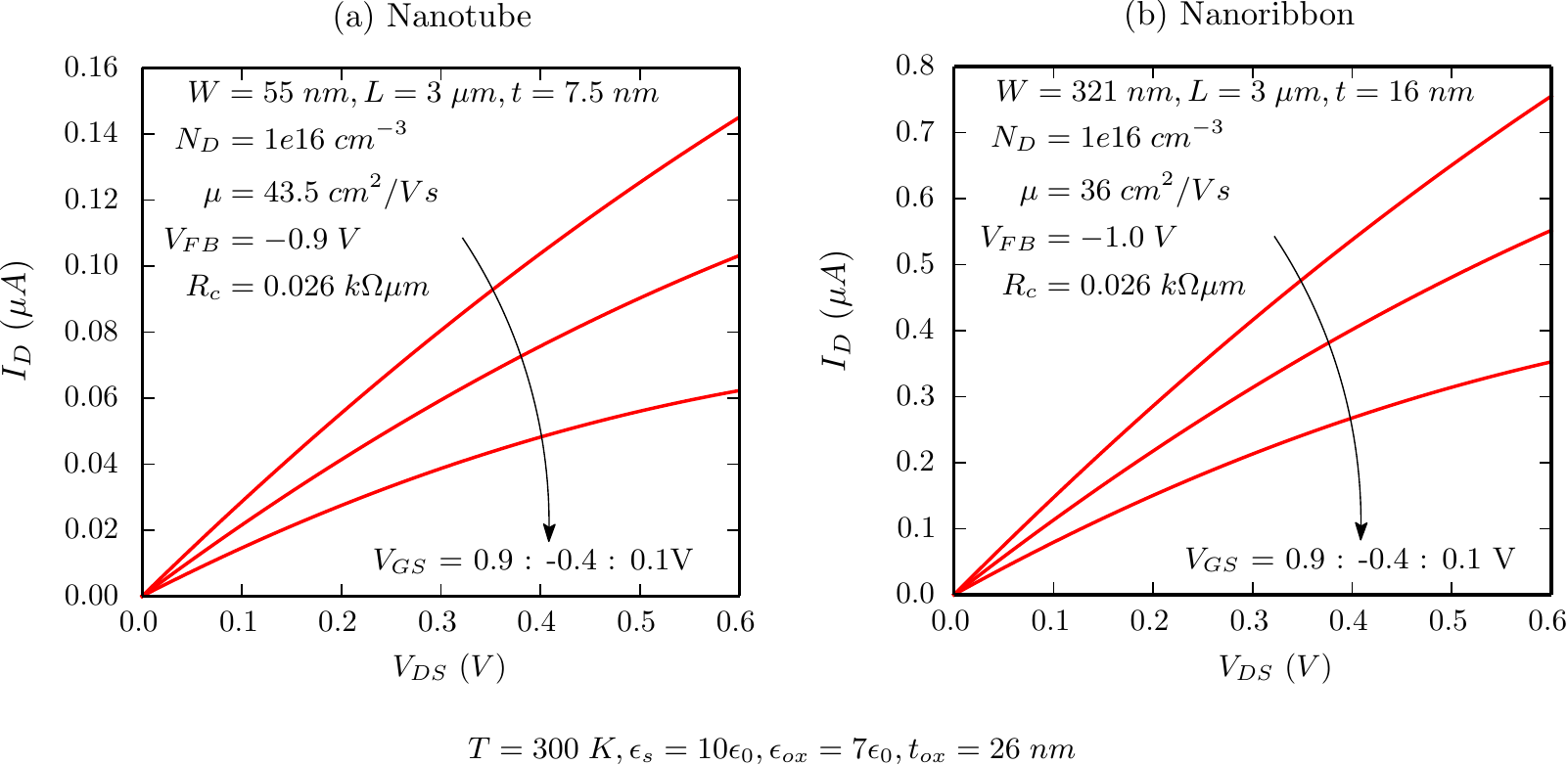}
\caption{Projected $I_D - V_{DS}$ characteristics of the NT and NR MOSFETs
assuming contact resistance $R_c = R_{c\;min} = 0.026 \; k\Omega \mu m$. Other
parameters are identical to those in Fig. \ref{fig_devicechar}. An improvement
of $\sim 27 \%$ ($\sim 36 \%$) is seen in the current at $V_{DS} = 0.6V$
for the NT (NR) as compared to Fig. \ref{fig_devicechar}.}
\label{fig_projchar}
\end{figure}

The  minimum  contact  resistance  to  a 2-D  material  is  given  by
$R_{c\;min} = 0.026 / \sqrt{n_s}$  $k\Omega \mu m$, where $n_s$ is the
sheet    charge    density    in    units    of    $1e13\;    cm^{-2}$
\cite{DJ_NatureMat_2014}.   Assuming  that  $n_s  = 1$  in  the  above
expression, the minimum contact  resistance is $R_{c\;min} = 0.026 \;
k\Omega \mu m$.  Note that  the value of contact resistance measured
in   our  NT  MOSFET   is   $R_c  =   39.6   \;  k\Omega   \mu
m$.  Fig.  \ref{fig_projchar}  predicts  the currents  that  can  be
obtained from our NT and  NR MOSFETs, provided the contact resistance
is chosen  to be $R_{c\;min}$.  Improving the contacts in  this manner
causes a $\sim 27 \%$ (  $\sim 36 \%$ ) increase in the currents
of the NT (NR) at $V_{DS} = 0.6V$. 
 
\section{Conclusion}
In conclusion,  we have modeled  the $I_D-V_D$ characteristics  of the
fabricated NT  and NR MOSFETs operating  in a regime  where the entire
channel is accumulated.   The results of this model  compare well with
the  experimental data. We  extract  effective mobilities  of $43.5\;
cm^2/Vs$ and $36\; cm^2/Vs$ for the NT and NR MOSFETs respectively.

\small{
\begin{singlespace}

\end{singlespace}

\end{document}